\documentclass[twocolumn,aps,prd]{revtex4}
\usepackage{amsmath,graphicx,bm}
\begin{document}

\title{Distinguishing Modified Gravity from Dark Energy}

\author{Edmund Bertschinger and Phillip Zukin}
\affiliation{Department of Physics, MIT, 77 Massachusetts Ave.,
Cambridge, MA 02139}

\begin{abstract}
The acceleration of the universe can be explained either through
dark energy or through the modification of gravity on large scales.
In this paper we investigate modified gravity models and compare
their observable predictions with dark energy models. Modifications
of general relativity are expected to be scale-independent on
super-horizon scales and scale-dependent on sub-horizon scales. For
scale-independent modifications, utilizing the conservation of the
curvature scalar and a parameterized post-Newtonian formulation of
cosmological perturbations, we derive results for large scale
structure growth, weak gravitational lensing, and cosmic microwave
background anisotropy. For scale-dependent modifications, inspired
by recent $f(R)$ theories we introduce a parameterization
for the gravitational coupling $G$ and the post-Newtonian parameter
$\gamma$. These parameterizations provide a convenient formalism for
testing general relativity. However, we find that if dark energy is
generalized to include both entropy and shear stress perturbations,
and the dynamics of dark energy is unknown a priori, then modified
gravity cannot in general be distinguished from dark energy using
cosmological linear perturbations.
\end{abstract}


\maketitle

\section{Introduction}
\label{sec:intro}

Cosmic acceleration has been revealed by measurements of the
redshift-distance relation $\chi(z)$ where $\chi$ is the comoving
radial distance.  The Hubble expansion rate follows from
$H(z)=(d\chi/dz)^{-1}$ (in units where $c=1$). This determination
assumes only that the observable universe is adequately described by
the Robertson-Walker metric, an assertion that is testable
empirically \cite{Hogg04,LuHellaby07} and does not imply the
validity of general relativity (GR).

The inference of dark energy follows once the Einstein field
equations of general relativity are imposed on the Robertson-Walker
metric yielding the Friedmann equations. These equations imply that
a stress-energy-momentum component with negative pressure is needed
to explain cosmic acceleration. This substance may be vacuum energy
(i.e., a cosmological constant, giving rise to the $\Lambda$CDM
model) or a scalar field \cite{Copeland06}. The dark energy equation
of state for uniform expansion, $p(\rho)$, can be determined from
measurement of $\rho(z)$, which itself follows from $H(z)$ combined
with the first Friedmann equation. Measuring $w(z)=p/\rho$ is the
primary goal of dark energy experiments \cite{DETF}.

Another possibility is that general relativity requires modification
on large distance scales and at late times in the universe.  In this
case cosmic acceleration would arise not from dark energy as a
substance but rather from the dynamics of modified gravity. Modified
gravity is not particularly attractive theoretically, but the
observed cosmic acceleration is so surprising that all plausible
explanations should be considered.

Observations of the cosmic expansion history cannot distinguish dark
energy from modified gravity \cite{fRexphist,Song06}.  Testing
gravity requires exploring the evolution of spatial inhomogeneity
(e.g.\ \cite{ZLBD07, inhom} and references therein).  Modified gravity
theories must pass tests within the solar system and in relativistic
binaries \cite{Will06}.  They are expected to show significant
departures from general relativity only on cosmological distance
scales. The combination of cosmic microwave background anisotropy,
weak gravitational lensing, and the growth of clustering of dark
matter and galaxies provides an opportunity to discriminate between
dark energy and modified gravity.

Performing cosmological tests of modified gravity requires a set of
predictions.  There are two approaches to generating these
predictions. In the first, a theory is specified by its Lagrangian
(or other fundamental description) which provides the equations of
motion for both homogeneous expansion and cosmological
perturbations.  A class of theories can be specified by giving a
Lagrangian with free parameters, e.g. $f(R)$ theories where $R$ is
the Ricci scalar \cite{Song06}.

A second approach is inspired by the parameterized post-Newtonian
framework for solar system tests \cite{ThorneWill71}. Here one
begins with the solution of the gravitational field equations (i.e.,
the metric) instead of the Lagrangian.  Several authors have
recently adopted this framework or a similar one
\cite{HuSawicki07b,Caldwell07,Amendola07,Amin07,Linder07}. The difficulty here is to find
a good ``Newtonian'' description in cosmology to which one adds
``Post-Newtonian'' parameters. Metric perturbations in the scalar
sector governing the growth of cosmic structure are characterized by
two spatial scalar fields, $\Phi$ (the Newtonian potential) and
$\Psi$ (the spatial curvature potential). Even if we introduce the
relationship $\Psi=\gamma\Phi$ with Eddington parameter $\gamma$
\cite{Eddington22}, there remains one unknown function of space and
time. In the solar system case, by contrast, $\Phi=-GM/r$ is known
to provide an excellent approximation to planetary dynamics.

On solar system scales, and even within galaxies, one can use
test-particle orbits to determine $\Phi$, and light rays to
determine $\Phi+\Psi$.  This comparison yields impressive limits on
$|\gamma-1|$ in the solar system \cite{Will06}.  However, on a scale
of several kpc, gravitational lensing combined with stellar dynamics
in elliptical galaxies yields a current best result
$|\gamma-1|=0.02\pm0.07$ \cite{Bolton06}.  At the scale of Gpc where
dark energy appears to drive accelerated expansion, there are no
longer any bound test-particle orbits to measure $\Phi$, so a
different approach, based on cosmological perturbation theory, is
needed.

Previous work in the cosmological parameterized post-Newtonian
framework has either assumed that some of the Einstein field
equations remain valid with modified gravity \cite{Caldwell07} or
has examined the dynamics of individual theories, e.g.\
\cite{Dore07}.  Neither approach is ideal. One would prefer to
sample all possible theories in a broad class, and for each theory
to constrain the potentials by a consistency condition that does not
assume general relativity or any particular modification thereof. 

Such a consistency condition was found recently in Ref.\
\cite{Bertschinger06} for the long-wavelength perturbations of a
Robertson-Walker spacetime.  This result was derived assuming that
gravity is described by a classical four-dimensional metric theory
having a well-defined infrared limit (i.e., the theory is well
behaved for very long wavelength perturbations).  For practical
application, assumptions must also be made about the background
spatial curvature and entropy perturbations.  Assuming an
inflationary or equivalent origin of perturbations, long-wavelength
isentropic perturbations are imprinted in the spatial curvature on a
flat background.  In general relativity, these spatial curvature
fluctuations, represented by the gauge-invariant $\zeta$ field of
Bardeen et al.\ \cite{BST83} or the ${\cal R}$ field of Lyth
\cite{Lyth85}, are time-independent in the long-wavelength limit.
Ref.\ \cite{Bertschinger06} presented a derivation of the conserved
curvature perturbation (calling it $\kappa$) making no assumption
about the field equations except that they have a well-defined
infrared limit. Physically this means that curvature perturbations
are small and that all waves propagate causally.  In what follows,
the curvature perturbation introduced in Ref. \cite{Bertschinger06}
will be called $\zeta$ although its definition differs from that of
Ref. \cite{BST83}. For long wavelength perturbations on a flat
background, ${\cal R}=\zeta$.

In the long-wavelength limit all cosmological perturbations
factorize into functions of time multiplying the curvature
perturbation or its spatial derivatives.  This is true in GR and in
modified gravity theories that are well-behaved in the infrared
limit. One might naively expect this factorization to hold only on
scales larger than the Hubble length. In general relativity,
however, signals in the scalar sector propagate at the speed of
sound, not the speed of light \cite{Bertschinger96}, leading to
conservation of $\zeta$ on scales larger than the Jeans length.

To satisfy solar system tests, modified gravity theories for cosmic
acceleration must introduce a length scale $L_G$ below which general
relativity is recovered. This length scale might be associated, for
example, with the dynamics of new scalar degrees of freedom.  The
value of this length scale, compared with the size of the systems
investigated, plays a crucial role in characterizing the behavior of
modified gravity theories.

If $L_G$ is much smaller than the length scales over which linear
cosmological structure formation is measured (e.g., $L_G=1$ Mpc),
then the factorization of cosmological perturbations on scales
larger than the Jeans length remains valid.  We denote this case
scale-independent modified gravity.  These theories are like GR in
that the curvature perturbation is conserved for the relevant length
scales.  This condition yields a great simplification of the
dynamics, reducing cosmological perturbations to quadratures.

In GR, waves propagating in the scalar sector travel only at the
speed of sound, so that scale-dependence of transfer functions
arises only below the Jeans length.  Modified gravity theories,
however, typically have additional fields supporting waves that
travel at the speed of light.  In this case, $L_G$ is the Hubble
length and the factorization of cosmological perturbations no longer
holds.  Theories of this type are called scale-dependent modified
gravity theories.  Now two quantities, $\gamma$ and the
gravitational coupling $G_\Phi$ (the generalization of Newton's
constant in the Poisson equation for $\Phi$), are needed to
characterize gravity, and both will vary with length scale as well
as with time.  Even such complicated models can still be
approximated by parameterizations, as we will discuss below. Ref. \cite{HuSawicki07b} found a way of bridging super- and sub-horizon modifications to GR. Our parametrization, while not as general, is simpler because it only involves a few free parameters and no free functions. 

Because we do not start with a Lagrangian, we cannot explain cosmic
acceleration.  We take the cosmic expansion history as given from
observations. Rather than providing a complete theory of modified
gravity, we provide a framework for observational tests of gravity
in cosmology.

This paper is organized as follows.  Section \ref{sec:long}
describes the curvature perturbation and its use to build
scale-independent modified gravity theories.  Section
\ref{sec:observe} works out the growth of structure on sub-horizon
scales, shows that the Poisson equation is modified, and derives
results for cosmic microwave background anisotropy and weak
gravitational lensing for scale-independent modified gravity
theories.  Section \ref{sec:f(R)} then examines the sub-horizon
behavior of a currently popular class of theories known as $f(R)$
models and shows that they are scale-dependent.  Section
\ref{sec:shear} considers the alternative hypothesis that dark
energy has a peculiar stress tensor while gravity is governed by GR.
Finally, results are summarized and conclusions are presented in
Section \ref{sec:discuss}.


\section{Gravity at long wavelengths}
\label{sec:long}

Our starting point is the perturbed Robertson-Walker metric in
conformal Newtonian gauge \cite{MFB92}:
\begin{equation}\label{pertFRW}
  ds^2=a^2(t)[-(1+2\Phi)dt^2+(1-2\Psi)\gamma_{ij}dx^idx^i]\ ,
\end{equation}
where $t$ is conformal time, $a(t)=1/(1+z)$ is the expansion scale
factor, and $\gamma_{ij}({\bf x},K)$ is the three-metric for a space
of constant spatial curvature $K$, e.g.
$\gamma_{ij}dx^idx^j=d\chi^2+r^2(\chi,K)d\Omega^2$ where
$r(\chi,K)\sqrt{K}=\sin(\chi\sqrt{K})$ for $K>0$ and is analytically
continued for $K\le0$. Note that different conventions appear in the
literature for the metric perturbations: $\Phi=\Psi_{\rm Hu}=\psi$
and $\Psi=-\Phi_{\rm Hu}=\phi$ where $(\Psi_{\rm Hu},\Phi_{\rm Hu})$
are the potentials of Ref.\ \cite{HuSawicki07b} and $(\psi,\phi)$
are the potentials of Ref.\ \cite{MaBert95}. Linear perturbation
theory is assumed to be valid throughout this paper.

The evolution of the scale factor can depend, in principle, on any
quantities characterizing the geometry and composition of the
Robertson-Walker background, for example the spatial curvature $K$
and the entropy density (or equivalently, parameters characterizing
the equation of state).  We neglect entropy perturbations and
consider only curvature perturbations on a flat ($K=0$) background.
Assuming that the unknown gravity theory has a well-defined infrared
limit obeying causality, long wavelength curvature perturbations
should evolve like separate Robertson-Walker universes.  In this
case it is possible to transform to a new set of coordinates, $t\to
t-\alpha(t)$ and $\chi\to\chi(1+\zeta)$ where $\zeta$ is constant
and $\dot\alpha=\Phi+\Psi-\zeta$, such that the new line element is
eq.\ (\ref{pertFRW}) with $\Phi=\Psi=0$ and having spatial curvature
$K(1+2\zeta)$. Thus, $\zeta$ is one-half the spatial curvature
perturbation. Enforcing the coordinate transformation leads to the
consistency condition \cite{Bertschinger06}
\begin{equation}\label{consistent}
  \frac{1}{a^2}\frac{\partial}{\partial t}\left(\frac{a^2\Psi}
    {\cal H}\right)+\Phi-\Psi=\left[\frac{1}{a}\frac{\partial}
    {\partial t}\left(\frac{a}{\cal H}\right)+\frac{K}{{\cal H}^2}
    +O(k^2)\right]\zeta\ ,
\end{equation}
where ${\cal H}=\dot a/a=aH$ and $k$ is the comoving wavenumber.
Although Ref. \cite{Bertschinger06} states that large-scale shear
stress is neglected in this result, in fact eq.\ (\ref{consistent})
is valid for $k\to0$ in general relativity (and presumably in
modified gravity theories) even if shear stress is present.  The
curvature term $K/{\cal H}^2$ has been computed assuming the
Friedmann equation is valid; in modified gravity theories this term
might be different but it must vanish when $K=0$. Hereafter we
assume $K=0$ and drop the curvature term.

Eq.\ (\ref{consistent}) may be regarded as a definition of the
curvature perturbation $\zeta$ for arbitrary theories of gravity.
For long wavelengths, $\zeta$ is independent of time. Sound waves in
the matter sector or wave propagation in the modified gravity sector
cause $\zeta$ to change with time on small scales.  These changes
are implied by the neglected terms proportional to $k^2\zeta$ in
eq.\ (\ref{consistent}). For now we ignore such terms, in effect
assuming that both the Jeans length and $L_G$ are smaller than the
cosmological scales of interest.

During the radiation-dominated era the Jeans length is comparable to
the Hubble length, and $\zeta$ (and $\Phi$ and $\Psi$) is damped for
scales smaller than the Jeans length.  We assume that during this
early period of evolution general relativity is an excellent
approximation so that the damping is well described by the transfer
functions computed using standard codes \cite{MaBert95,cmbfast}.
When modified gravity becomes important at low redshift, the Jeans
length has dropped to a few Mpc or less.  In practice, we modify
CMBFAST only for $z<30$ and then do so in such a way as to enforce
eq.\ (\ref{consistent}) with $\zeta$ corrected from its primeval
value using the GR transfer function at $z=30$.

Assuming a well-defined infrared limit, the time and space
dependence of perturbations must factorize for wavelengths longer
than the Jeans length or $L_G$, e.g.
\begin{equation}\label{factorize}
  \Phi({\bf k},t)=F(a)\zeta({\bf k})+O(k^2\zeta)
\end{equation}
where ${\bf k}$ is the wavevector.  Factorization implies that the
ratio of the two gravitational potentials depends only on time as
$k\to0$.  Therefore we may write, for any causal theory of gravity
having a well-defined infrared limit,
\begin{equation}\label{gammadef}
  \Psi({\bf k},t)=\gamma(a)\Phi({\bf k},t)+O(k^2\zeta)\ .
\end{equation}
In modified gravity theories, $\gamma(a)$ is the only degree of
freedom important for long-wavelength scalar perturbations.  The
conditions of causality and a well-defined infrared limit greatly
restrict the dynamics of modified gravity theories.

We now make the key assumption that the terms proportional to
$k^2\zeta$ in eqs.\ (\ref{consistent})--(\ref{gammadef}) can be
neglected not only on super-horizon scales $k<{\cal H}^{-1}$ but
also, as they can be in general relativity, on sub-horizon scales
down to the Jeans length.  This assumption defines a class of
theories we call scale-independent modified gravity models.

Under these assumptions, modified gravity is completely specified on
large scales by the scale-independent function $\gamma(a)$. At high
redshift when dark energy is unimportant we require $\gamma\to1$ in
order to retain the success of general relativity in explaining the
cosmic microwave background anisotropy \cite{Spergel07}.  Thus we
adopt the following parameterization for scale-independent modified
gravity:
\begin{equation}\label{gamma}
  \gamma(a) = 1+\beta a^s\ ,
\end{equation}
where $\beta$ and $s>0$ are constants.  Eqs.\
(\ref{consistent})--(\ref{gamma}) now give
\begin{equation}\label{Fquad}
  \gamma F(a)=a^{-2}{\cal H}\gamma^{(1/s)}
    \int^a_0 a\gamma^{-(1/s)}\frac{d}{da}
    \left(\frac{a}{\cal H}\right)\,da\ .
\end{equation}
Changing the lower limit of integration introduces a rapidly
decaying solution which we ignore.

In general relativity with negligible shear stress, $\gamma=1$. When
a component with constant equation of state parameter
$w>-\frac{1}{3}$ is dominant, $a\propto t^n$ with $n=2/(1+3w)$
yielding $F=(3+3w)/(5+3w)$.  Thus, for long wavelengths the
potential drops from $\Phi=\frac{2}{3}\zeta$ during the
radiation-dominated era to $\Phi=\frac{3}{5}\zeta$ during the
matter-dominated era.

\begin{figure}[t]
  \begin{center}
    \includegraphics[height = 70mm, width=90mm]{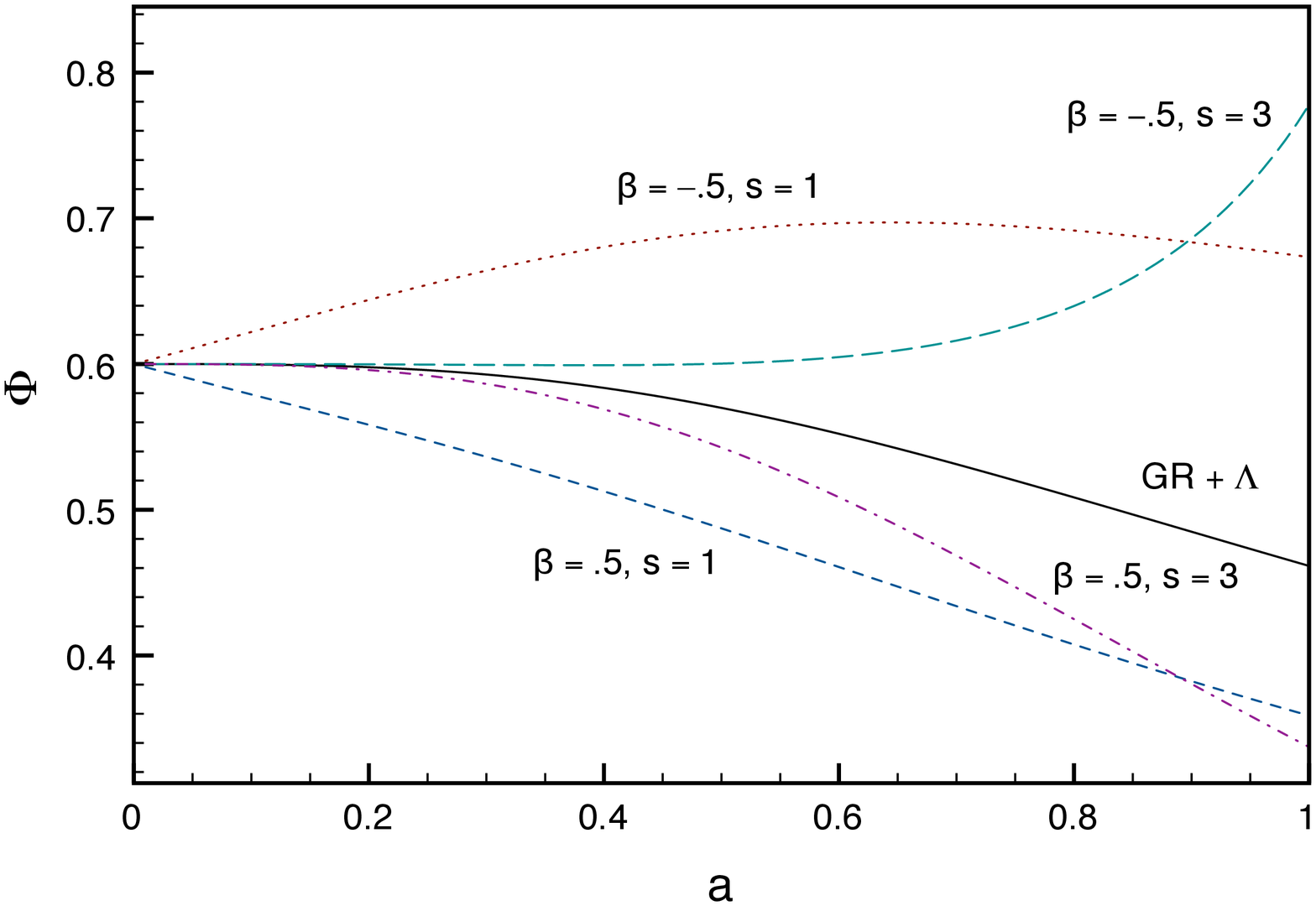}
    \includegraphics[height = 70mm, width=90mm]{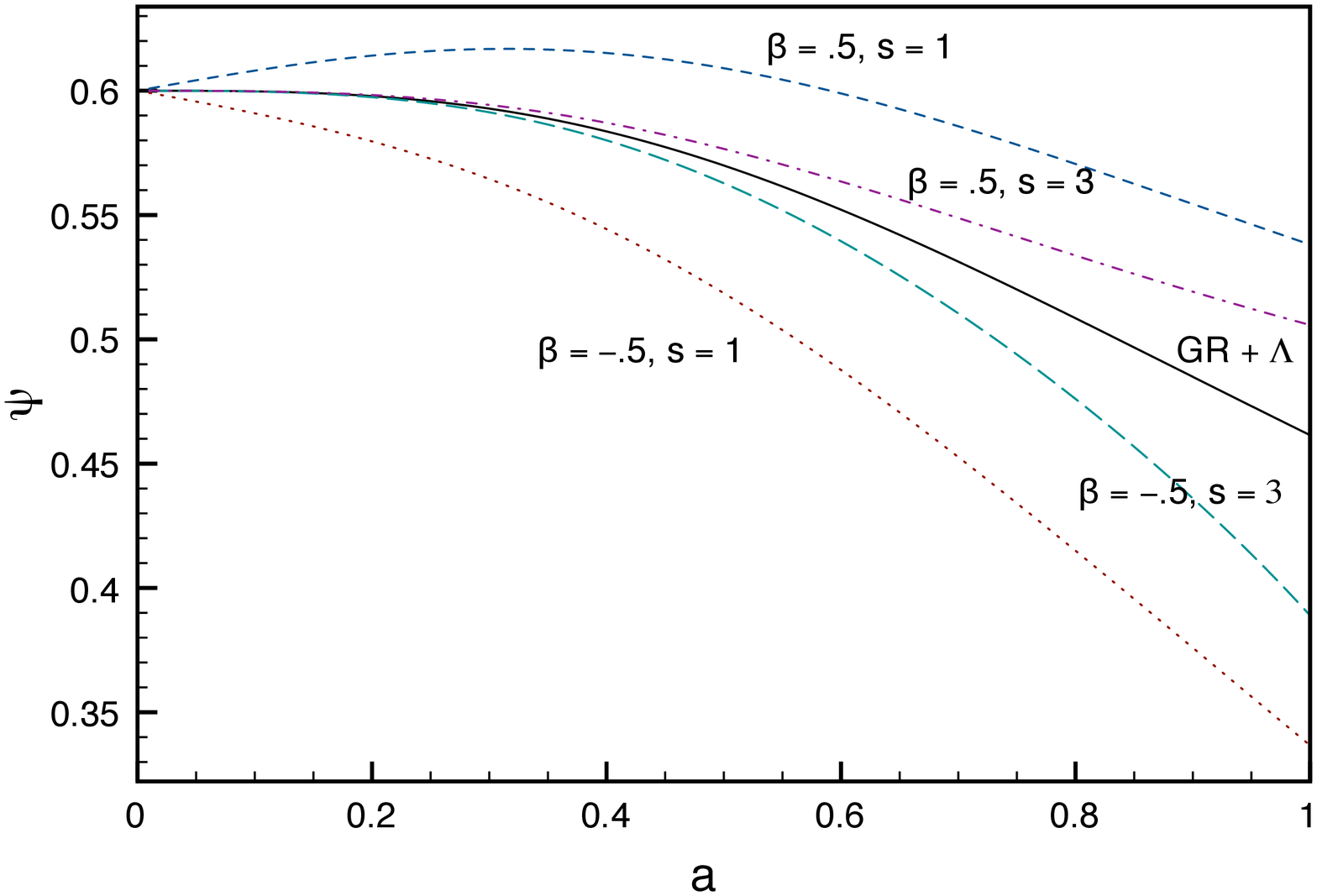}
  \end{center}
\caption{Evolution of the scalar potentials $\Phi$ and $\Psi$ in the
long wavelength, scale-independent limit, assuming a $\zeta=1$
normalization. Modified gravity effects, parameterized by eq.\
(\ref{gamma}), arise later for larger $s$.  The GR model assumes
$\gamma=1$ and a cosmological constant. For $\gamma<1$ the Newtonian
potential grows, unlike general relativity with a cosmological
constant.} \label{fig:F}
\end{figure}

Of greater interest here is the evolution of the potentials during
the matter-dominated era with modified gravity parameterized by
(\ref{gamma}). Figure \ref{fig:F} shows the results for $s=1$ and
$s=3$ as well as the GR case $\beta=0$.  The background expansion
history is chosen to match GR with $\Omega_m=0.284$ and a
cosmological constant with $\Omega_\Lambda=0.716$. The choice $s=3$
matches Caldwell et al.\ in the limit $\beta\ll 1$
\cite{Caldwell07}. However, as we will see below, our results differ
from theirs because they assumed the validity of some components of
the Einstein field equations, while we instead required consistent
causal evolution on large scales. This difference will be discussed
further below in Section \ref{sec:shear}.

Figure \ref{fig:F} shows that the Newtonian potential $\Phi$ is
enhanced and the spatial curvature $\Psi=\gamma\Phi$ is diminished
for $\gamma<1$, compared with general relativity.  For $s=3$ the
modifications occur later because $|\gamma-1|$ is smaller at earlier
times.  The quantitative results depend on the validity of eq.\
(\ref{consistent}) but this qualitative behavior (the Newtonian
potential being enhanced for $\gamma<1$) should persist in general.


\section{Observables for scale-independent modified gravity}
\label{sec:observe}

With the time evolution of the metric in hand for long wavelengths
we are now able to calculate observable quantities for
scale-independent modified gravity theories parameterized by the
constants $(\beta,s)$. The effects considered here are the growth of
structure in the dark matter, microwave background anisotropy, and
weak gravitational lensing.

\subsection{Growth of structure}

Until now, no assumptions have been made about dynamics in the
matter sector except for causality and consistency with a spatially
homogeneous, uniformly expanding Robertson-Walker solution. To
follow the growth of structure we must specify how the matter fields
are coupled to gravity.  Here we assume that the dark matter obeys
the weak equivalence principle, i.e.\ collisionless dark matter
particles follow geodesics.  This choice explicitly forces
scalar-tensor theories to the Jordan frame in which matter fields
are minimally coupled to gravity. 

In the conformal Newtonian gauge, on sub-horizon scales where
$|\delta|\gg|\Psi|$ with $\delta\equiv\delta\rho/\rho_m$, and
$\rho_m$ is the average mass density, cold dark matter fluctuations
obey the evolution equation
\begin{equation}\label{cdmevol}
  \ddot\delta+{\cal H}\dot\delta=-k^2\Phi\ .
\end{equation}
This equation follows from particle number conservation and geodesic
motion or, equivalently, from energy-momentum conservation.  The
density perturbation field can be written as
\begin{equation}\label{Dfactor}
  \delta({\bf k},t)=-k^2D(a,k)\zeta_i({\bf k})
\end{equation}
where $\zeta_i$ is the curvature perturbation at $a=a_i$ which we
take to be $z=30$ so that the Jeans length is smaller than the
scales of interest and modified gravity has not yet become
important. For $a>a_i$, $\Phi$ factorizes and eq.\ (\ref{cdmevol})
can be reduced to quadratures for $D(a,k)$. With initial conditions
$D(a,k)=D_i(k)$ and $\partial_t D(a,k)=\dot D_i(k)$ at $a=a_i$, the
solution is
\begin{equation}\label{Density}
  D(a,k)=D(a)+D_i(k)+a_iy(a)\dot D_i(k)\ ,
\end{equation}
where
\begin{subequations}\label{Dysol}
\begin{eqnarray}
  D(a)&\equiv&y\int^a_{a_i}\frac{F}{\cal H}\,da -
    \int^a_{a_i}\frac{yF}{\cal H}\,da\ ,\label{Dsol}\\
  y(a)&\equiv&\int^a_{a_i}{\frac{da}{a^2{\cal H}}}\ .
    \label{ydef}
\end{eqnarray}
\end{subequations}
The function $y(a)$ asymptotes to a constant but $D(a)\propto a$ for
$a\gg a_i$ in the matter-dominated era.  Thus the late-time solution
for density perturbation growth factorizes, $D(a,k)=D(a)$ in eq.\
(\ref{Dfactor}).  This perturbation growth is often represented as a
function of redshift by defining $g(z)\equiv D(a)/D(1)$ with
$a=1/(1+z)$.

\begin{figure}[t]
  \begin{center}
    \includegraphics[height = 70mm, width=90mm]{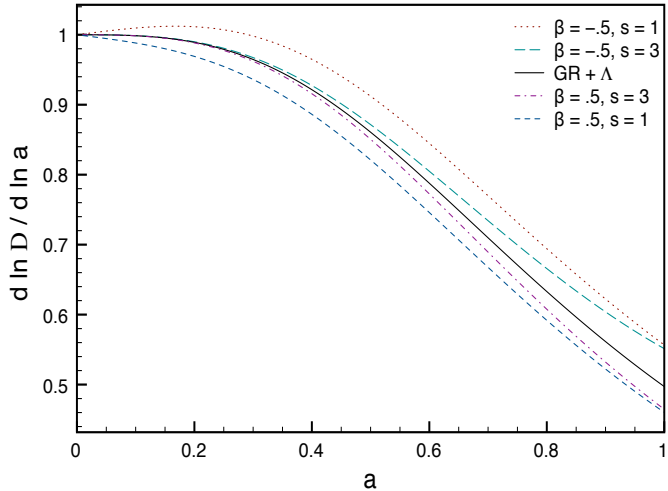}
  \end{center}
\caption{Evolution of the logarithmic density perturbation growth
rate $d\ln D/d\ln a$.  Models with $\gamma<1$ have enhanced growth
relative to the $\Lambda$CDM (GR) model.} \label{fig:I}
\end{figure}

Figure \ref{fig:I} shows the logarithmic derivative of the density
perturbation growth versus time for our parameterized modified
gravity models.  In the $\Lambda$CDM model, $d\ln D/d\ln
a\approx[\Omega_m(a)]^{6/11}$ \cite{Nesseris07}. The transition to a
cosmological constant-dominated universe leads to a suppression of
growth.  If instead gravity is modified, the growth rate can be
increased or decreased relative to the GR case. The qualitative
effects are easy to understand.  We already saw that models with
$\gamma<1$ ($\beta<0$) get an enhanced Newtonian potential $\Phi$. A
stronger potential increases the gravitational force on density
perturbations leading to an enhanced growth rate.

The simplest test of growth of perturbations is the total
perturbation growth by redshift zero, which is characterized by the
variance of density fluctuations in spheres of radius
$R_8=8\,h^{-1}$ Mpc,
\begin{equation}\label{sigma8}
  \sigma^2_8=\int^{\infty}_0\frac{d^3k}{(2\pi)^3}P_m(k)W^2(kR_8)\ ,
\end{equation}
where $W(x)=3j_1(x)/x$.  The power spectrum of matter density
fluctuations is
\begin{equation}\label{powerspec}
  P_m(k)=\frac{2\pi^2}{k^3}\Delta_{\cal R}^2\left(\frac{k}{k_0}\right)^{n_s-1}
    T^2_m(k,z=30)\frac{D^2(z=0)}{D^2(z=30)}\ ,
\end{equation}
where $\Delta_{\cal R}$ is the amplitude of the initial scalar
curvature fluctuations on scale $k_0$ and $T_m$ is the transfer
function for matter fluctuations in the synchronous gauge relative to $\zeta = {\cal{R}}$ computed by CMBFAST (which accounts for the
suppression of growth during the radiation-dominated era). We adopt
$\Delta_{\cal R}^2=2.4\times 10^{-9}$, $k_0=0.002$ Mpc$^{-1}$, and
$n_s=0.958$ \cite{Spergel07}.

Our modification of gravity is scale-invariant in that the
$k$-dependence of the dark matter power spectrum is unchanged
relative to GR. Thus the amplitude of density perturbations
$\sigma_8$ depends on modified gravity only through the enhancement
or diminution of growth shown in Fig.\ \ref{fig:I}.

The specific results obtained here assume that the gravitational
potentials factorize on scales larger than a few Mpc. If this is
true, then CMB and galaxy clustering results on all scales can be
fit by a single modified gravity theory with one function
$\gamma(a)$ or, equivalently, $D(a)$.  If, however, modified gravity
introduces a length scale $L_G$ intermediate between $R_8$ and the
Hubble length, then the growth of structure will depend on
wavenumber in a way not described by a scale-independent $D(a)$. In
Section \ref{sec:scaledep} below we consider an alternative
scale-dependent parameterization of modified gravity and examine the
observable consequences.  For now, we assume $L_G<R_8$ or
equivalently that super-horizon relations apply, as they do in GR,
to sub-horizon scales all the way down to the larger of the Jeans
and nonlinear lengths.

\begin{figure}[t]
  \begin{center}
    \includegraphics[height = 80mm, width=100mm]{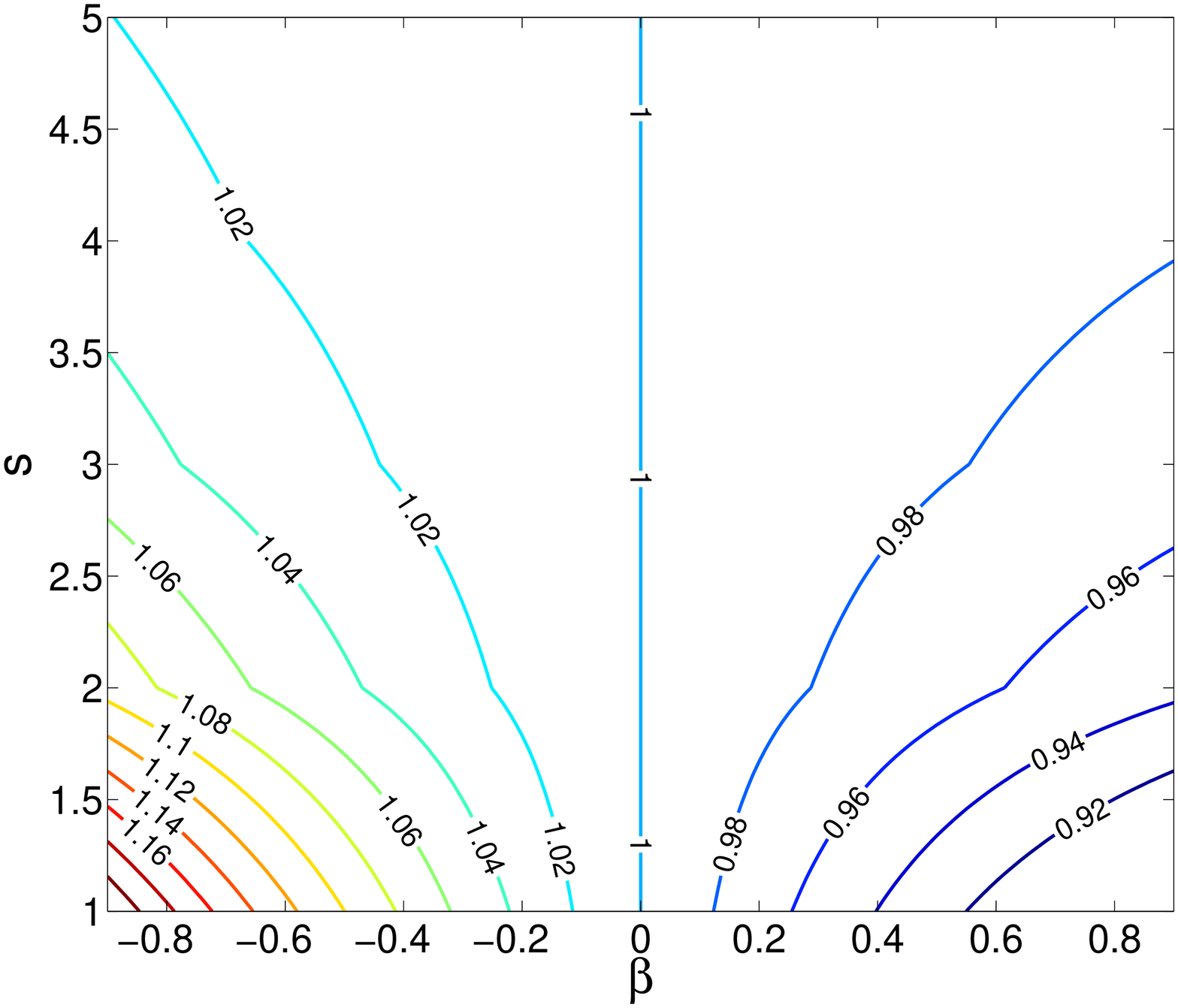}
  \end{center}
\caption{Contour plot of $\sigma_8(\beta,s)/\sigma_8(\Lambda {\rm
CDM})$ with $\beta$ running along the horizontal axis and $s$
running along the vertical axis.} \label{fig:Sigma}
\end{figure}

Figure \ref{fig:Sigma} shows a contour plot of $\sigma_8$
(normalized to the GR case) for different choices of the modified
gravity parameters.  As expected,
smaller values of $\gamma$ (i.e., $\beta<0$) lead to larger
amplitude.  Thus, modified gravity changes the amplitude of galaxy
clustering relative to the CMB, and could explain any apparent
discrepancy between the values of $\sigma_8$ inferred from CMB
analysis and galaxy clustering or lensing measurements.

\subsection{Modified Poisson Equation}

\begin{figure}[htb]
  \begin{center}
    \includegraphics[height = 70mm, width=90mm]{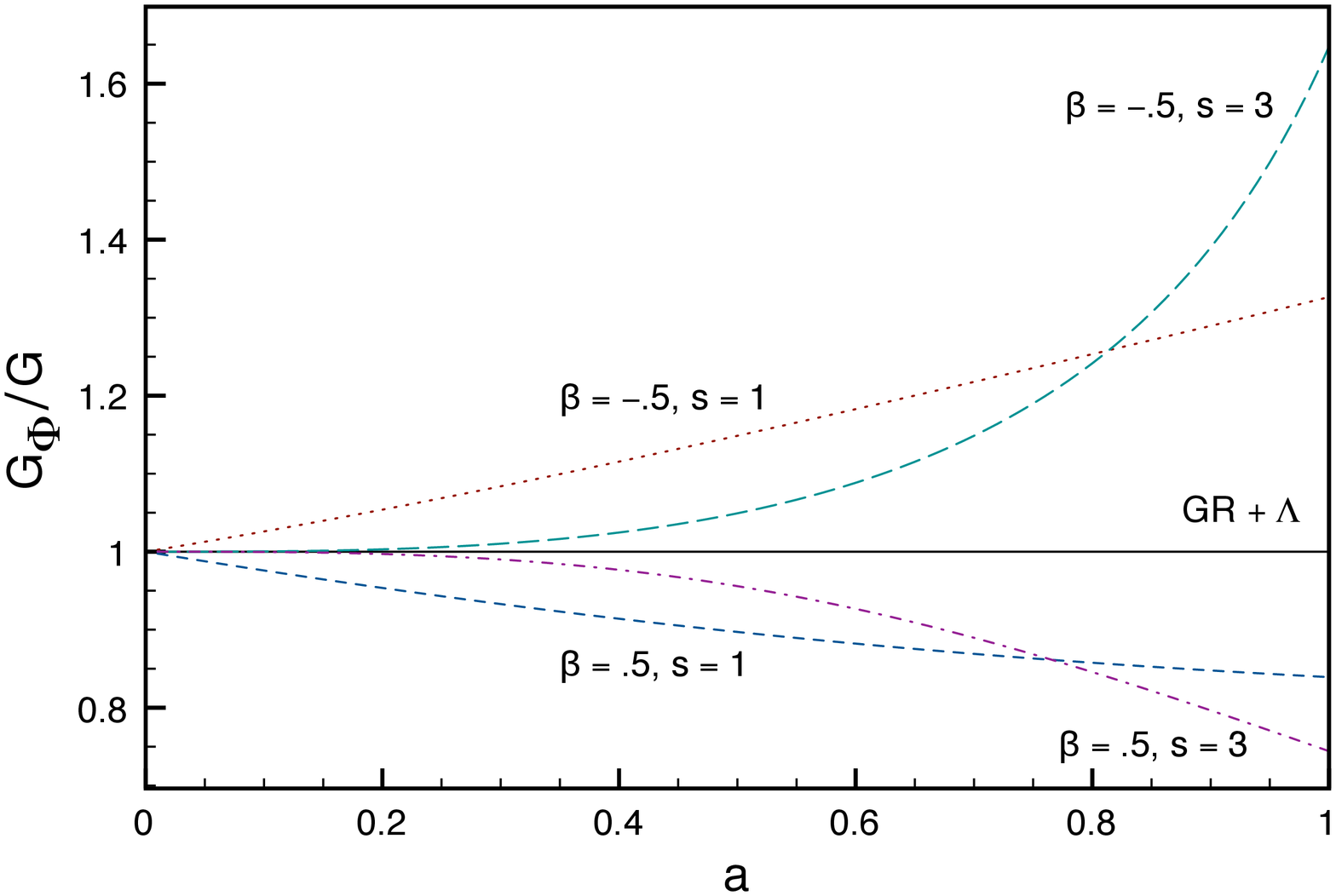}
    \includegraphics[height = 70mm, width=90mm]{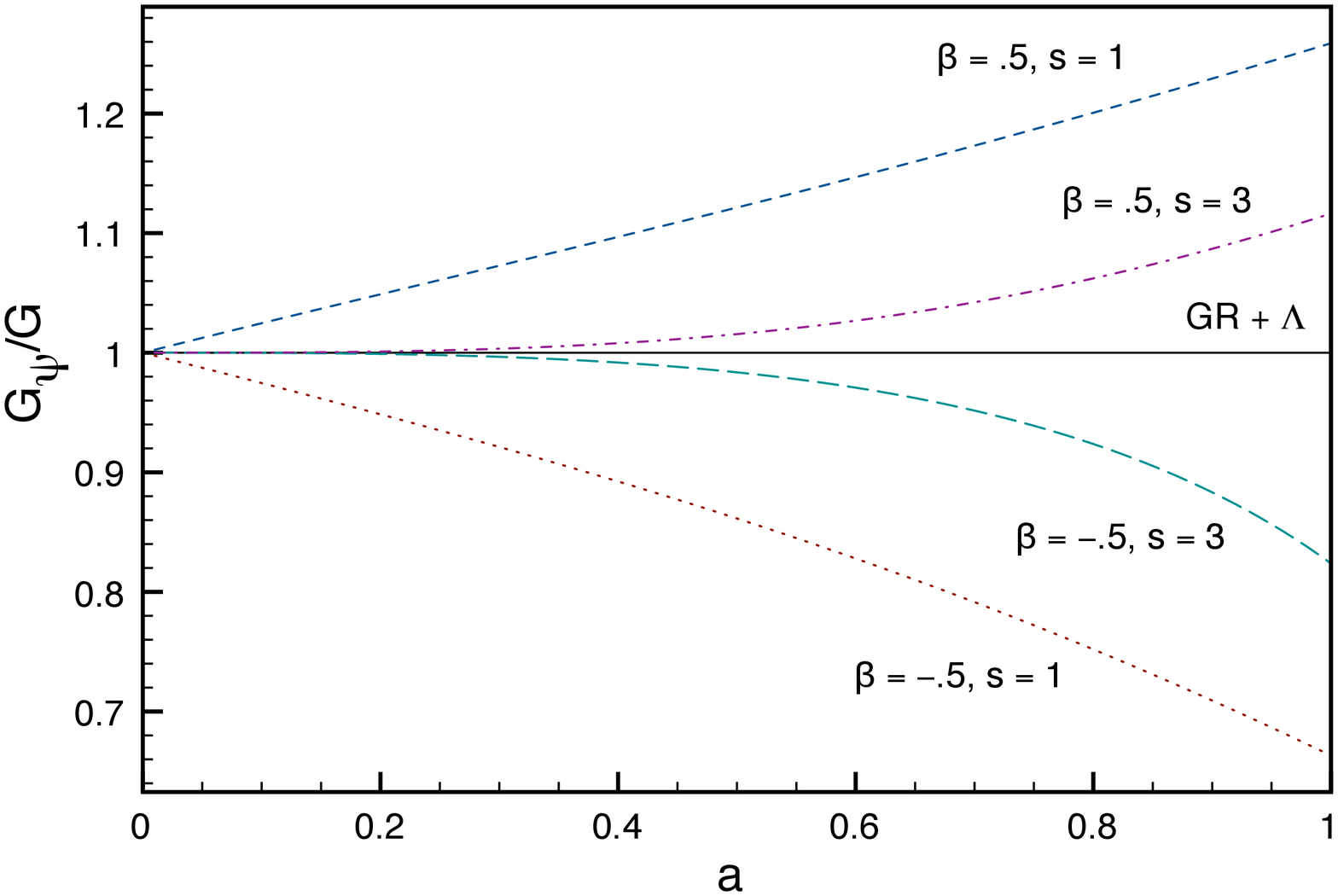}
  \end{center}
\caption{Evolution of $G_{\Phi}$ and $G_{\Psi}$. The behavior is
dominated by the potentials and exhibits the same features shown in
Fig.\ \ref{fig:F}.} \label{fig:G}
\end{figure}

Rearranging eqs.\ (\ref{factorize}) and (\ref{Dfactor}), we arrive
at a modified Poisson equation relating the Newtonian potential to
the overdensity:
\begin{equation}\label{Poisson}
  \nabla^2\Phi = \frac{F(a)}{D(a)}\delta \equiv 4\pi G_{\Phi}(a)
    a^2 \rho_m \delta\ .
\end{equation}
The space curvature potential obeys a similar Poisson equation,
\begin{equation}\label{Poisson-Psi}
  \nabla^2\Psi = 4\pi G_{\Psi}(a) a^2 \rho_m \delta\ ,
\end{equation}
where $G_{\Psi}(a) = \gamma G_{\Phi}(a)$. Plots of $G_{\Phi}$ and
$G_{\Psi}$ are shown in Figure \ref{fig:G}. Their time dependence is
dominated by the potentials $F(a)$ and $\gamma(a)F(a)$ since
$\delta$ is less sensitive to our modified gravity parameters
($\beta,s$). As a result, we see the same qualitative behavior as in
Fig.\ \ref{fig:F}. Models with $\gamma < 1$ produces a greater value
of $G_{\Phi}$ relative to the $\Lambda$CDM model, and larger values
of $s$ produce larger late time behavior.

In GR, the gravitational coupling $G$ is constant.  In many
alternative theories of gravity, the strength of gravity varies with
time (and also with place, for length scales smaller than $L_G$).
Time-varying $G$ is well known for scalar-tensor theories
\cite{Acquaviva05,MotaEmail}, but we find that it is a generic feature of all
modified gravity theories with $\gamma\ne1$ on cosmological scales.

The time-variation of $G$, represented by $\dot G/G$, has been
severely constrained by measurements in the solar system, in stars,
and in the early universe \cite{TestG}.  Limits on larger scales are
provided by the microwave background \cite{TestGCMB}.  The CMB
acoustic peak structure will be unaffected if variations occur only
long after recombination.  Modified gravity explanations of cosmic
acceleration suggest the need to constrain $\dot G/G$ on large
length scales in the late universe. It would be very interesting to
know, for example, to what extent the structure of clusters of
galaxies and their X-ray emission can be used to constrain $\dot
G/G$.

Note that the method used to derive our modified Poisson equations
is roundabout.  Had we started with a Lagrangian, the gravitational
field equations would directly yield the gravitational coupling
strength.  Because we started with a phenomenological description of
modified gravity, we instead deduce the dynamics of this coupling
from the requirements of causal evolution and the weak equivalence
principle.  On small scales our treatment breaks down as
modifications of general relativity must become scale-dependent.
Nonetheless, the motivation to investigate limits on $\dot G/G$ on
scales much larger than the solar system remains valid.

\subsection{CMB temperature anisotropy}

Modified gravity (or dark energy) affect the microwave background
only at late times through the integrated Sachs-Wolfe (ISW) effect:
\begin{equation}
  \frac{\Delta T}{T}(\hat{n})= \int (\dot{\Psi} + \dot{\Phi})\,d\chi
\end{equation}
where $\chi$ is the comoving radius and $\hat n$ is the photon
direction.  The ISW effect arises when the gravitational potentials
change with time, as occurs during transitional periods in cosmic
evolution.  One such contribution occurs during the transition from
radiation to matter domination. The other contribution is occurring
today during the current transition to an accelerating expansion.
The physics governing the matter-radiation transition is well
explained by GR, while the physics governing the transition today is
(for the models considered here) dependent on modified gravity
parameters. Because the recent ISW effect arises relatively nearby
($z<2$), it shows up only at large angular scales.  The ISW
contribution from some modified gravity models has been studied in
several recent papers \cite{Song06,Caldwell07}.

\begin{figure}[t]
  \begin{center}
    \includegraphics[height = 70mm, width=90mm]{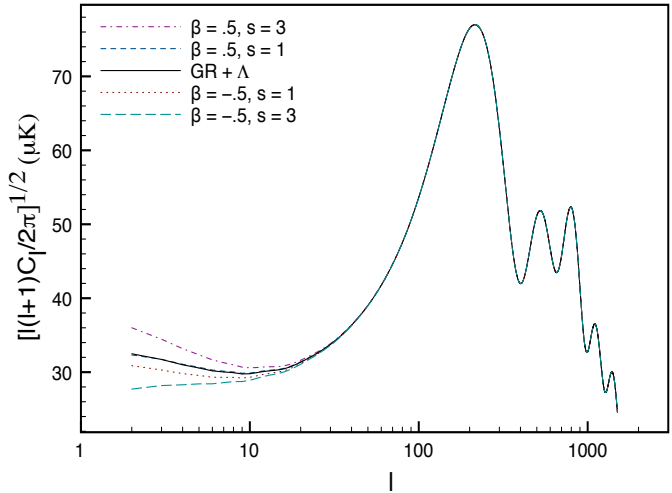}
  \end{center}
\caption{The temperature anisotropy power spectrum for different
parameter choices of modified gravity.} \label{fig:T}
\end{figure}

We computed the CMB temperature anisotropy spectrum by modifying
CMBFAST \cite{cmbfast} to replace the ISW contribution of the
$\Lambda$CDM model with that for our models assuming the
factorization of the potential.  The results are shown in Figure
\ref{fig:T}.  As expected, only the low-order multipoles are
affected.  Models with higher $s$ produce larger changes because
they lead to larger time derivatives.  Models with $0.2<\gamma<1$
($-0.8<\beta<0$) produce less anisotropy because of destructive
interference between the ISW and primary anisotropy contributions.
Although decreasing the quadrupole moment improves agreement with
observations, little statistical weight can be given to this
conclusion because the modifications, at least for $0.2\le\gamma
\le1.5$, are smaller than cosmic variance. However, cross correlating the CMB with galaxy surveys could potentially be a more discriminating probe of modified gravity \cite{HoEtAl08,Cooray02}

It was recently found \cite{Daniel08} that our results are consistent with recipe R1 of Caldwell et
al. \cite{Caldwell07}. However, we expect differences from our Figure \ref{fig:T} for recipe R3 of \cite{Caldwell07} since $\zeta$ is not conserved on large scales in this scheme. This difference will be discussed below in Section
\ref{sec:shear}.

\subsection{Weak lensing}

Metric perturbations $\Phi+\Psi$ affect both the energy of photons
(ISW effect) and their direction of travel (gravitational lensing).
Gravitational lensing causes both magnification (or
de-magnification) and differential stretching (shear) of background
images. The correlation function or power spectrum of weak
gravitational lens shear is an observable measure of large-scale
structure.  The weak lensing power spectrum is given by
\cite{WeakLens}
\begin{equation}\label{shearPS}
 P^{\kappa}_{l}=\int^{\chi_\infty}_0d\chi\,
 {W^2(\chi)}\frac{l^4}{\chi^4}P_{\Psi+\Phi}(k=\frac{l}
 {\chi}, \chi)\ ,
\end{equation}
where
\begin{equation}
  \frac{P_{\Psi+\Phi}}{(1+\gamma)^2}
  =\frac{2\pi^2\triangle^2_{\cal{R}}}{k^3}
  \left(\frac{k}{k_0}\right)^{n_s-1}T_{\Phi}^2
  (k,z=30)\frac{F^2(z)}{F^2(z=30)}\ ,\ \
\end{equation}
and
\begin{equation}
  W(\chi) =  \int^{\chi_\infty}_\chi d\chi' \frac{\chi' -
    \chi}{\chi'} \eta(\chi')\ .
\end{equation}
Here $\chi_\infty$ is the comoving distance to $z=10$ (the results
change by a negligible amount if the maximum redshift lies anywhere
between $6\le z\le15$), $T_{\Phi}$ is the transfer function of the
Newtonian potential relative to $\zeta$ computed at $z=30$ using CMBFAST and
$\eta(\chi)$ is the radial distribution of sources, normalized with
$\int\eta(\chi)\,d\chi=1$. Note that these formulae assume a flat
space.  Our lensing analysis uses the source distribution
\begin{equation}
  \eta(z)\propto z^2\exp[{-(1.41z/z_{\rm med})}^{1.5}]
\end{equation}
with $z_{\rm med}=1.26$ \cite{Massey}.  This distribution
approximates the galaxy redshift distribution of the COSMOS survey,
if there were no clumping.

\begin{figure}[t]
  \begin{center}
    \includegraphics[height = 70mm, width=90mm]{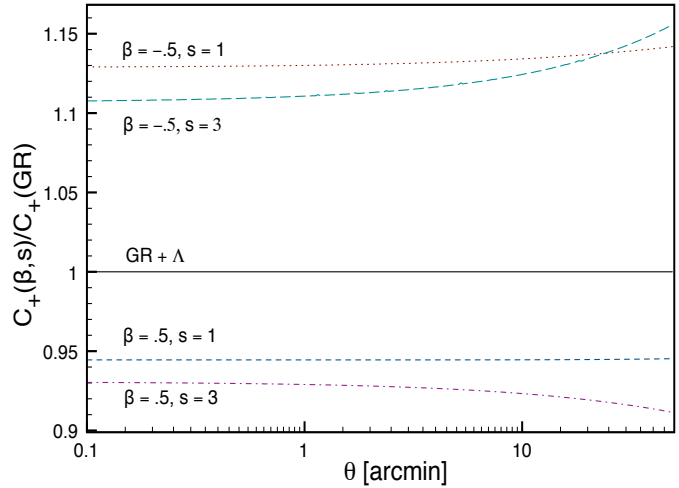}
  \end{center}
\caption{Ratios of weak lensing shear correlation $C_+$, to its
value for GR, for several sets of modified gravity parameters.}
\label{fig:C+}
\end{figure}

Measurements of weak gravitational lens shear, for galaxies
separated by angle $\theta$ on the sky, provide an estimate of shear
correlation functions including
\begin{equation}
  C_+(\theta)=\frac{1}{2\pi}\int^{\infty}_0 P^{\kappa}_{l}J_0(l\theta)
    l\,dl\ .
\end{equation}
Modifying gravity changes $F(z)$ thereby changing $C_+(\theta)$.
Figure \ref{fig:C+} plots $C_+(\theta)$ for modified gravity,
normalized at each $\theta$ by its GR value. At $z=1$, 1 Mpc corresponds to 2.15 arcmin. Hence, for some of the scales shown in Figure \ref{fig:C+} structures are nonlinear and thus beyond the regime of validity of the current framework. However, a scheme that takes a linear power spectrum to a nonlinear power spectrum would correct this flaw \cite{PeacockDodds}. On such small scales, our assumption of scale-independent modified gravity may also be invalid, so Figure \ref{fig:C+} should be regarded as suggestive, but not definitive, of modified gravity effects on weak lensing. 

As expected, models
with $\gamma<1$ have larger shear correlations because they have a
larger $F(z)$ and therefore more growth of structure (despite having
a smaller $1+\gamma$).  For angular scales less than about 10
arcmin, the effect is almost equivalent to a constant change in the
normalization of the power spectrum, i.e., in the value of
$\sigma_8$.  At larger angular scales, the redshift-dependence of
$F(z)$ at small redshift translates to a dependence on distance and
hence on angular scale, however this is in a regime where the shear
correlations are small and difficult to measure.  Thus,
scale-independent modified gravity theories predict an amplitude of
weak lensing different from GR with the same CMB primary anisotropy.
The acoustic peak amplitudes tightly constrain $\Delta_{\cal R}$. In
principle, measurements of $\sigma_8$ based on the CMB acoustic
peaks (which are unaffected by modified gravity) could differ both
from measurements based on galaxy clustering  [which depend on
$D(z)$] and those based on weak lensing [which depend on $F(z)$].
Current error bars are inconclusive \cite{Spergel07} but this
comparison of different $\sigma_8$ values could eventually provide a
powerful test of GR.

\section{Comparison with $f(R)$ theories}
\label{sec:f(R)}

Substantial work has already been done investigating modified
gravity effects for $f(R)$ theories.  Here we consider such theories
where the Ricci scalar $R$ is replaced by $R+f(R)$ in the
Einstein-Hilbert action, and where the action is extremized with
respect to the metric.  In these models, the field equations are
generically fourth-order.  In effect, modified gravity introduces a
new propagating scalar degree of freedom coupled to gravity, the
scalaron $f_R\equiv df/dR$ \cite{Starobinsky}.  The Compton
wavelength of the scalaron imprints a physical length scale, which
is made dimensionless by combining with the wavenumber $k$:
\begin{equation}\label{Qdef}
  Q\equiv\frac{3k^2}{a^2}\frac{f_{RR}}{1+f_R}\ ,
\end{equation}
where $f_{RR}\equiv d^2f/dR^2$.  Several papers have recently
discussed cosmological perturbation evolution for metric $f(R)$
theories \cite{f(R),HuSawicki07a,PS07}; our notation most closely
follows that of ref.\ \cite{PS07} except our potentials $\Phi$ and
$\Psi$ are exchanged from theirs.

In $f(R)$ theories, $\zeta$ is conserved on super-horizon scales
\cite{HuSawicki07b}.  However, the scalaron obeys a nonlinear
Klein-Gordon equation with two length scales: the Hubble length and
the scalaron Compton wavelength. The interesting case for
large-scale structure is the quasi-static regime of linear,
sub-horizon perturbations ($k^2\gg{\cal H}^2$) where
\cite{PS07}
\begin{equation}\label{fPoisson}
  \nabla^2\Phi\approx\frac{4\pi Ga^2\rho_m}{1+f_R}
    \left(\frac{3+4Q}{3+3Q}\right)\delta\ ,\ \
    \gamma\approx\frac{3+2Q}{3+4Q}\ .
\end{equation}
Differentiating eq.\ (\ref{consistent}) and substituting eqs.\
(\ref{cdmevol}) and (\ref{fPoisson}) along with the background
evolution equations (5) and (6) of ref.\ \cite{PS07} for a universe
containing only nonrelativistic matter and (optionally) a
cosmological constant yields
\begin{equation}\label{zetadot}
  \dot\zeta=U\dot\Psi+V\Psi\ ,
\end{equation}
where
\begin{equation}\label{Udef}
  U\equiv\frac{a{\cal H}}{\Gamma^2}\frac{\partial}{\partial t}
    \left(\frac{\Gamma^2B}{a{\cal H}^2}\right)+\frac{2QB}{3+2Q}
\end{equation}
and
\begin{eqnarray}\label{Vdef}
  V\equiv\frac{4\pi Ga^2\rho_m\Gamma B}{\gamma
   {\cal H}}+\frac{\partial}{\partial t}\left(\frac{B}{\gamma}
   \right)
   -\frac{\Gamma B}{{\cal H}a^2}\frac{\partial}{\partial t}
    \left[a\frac{\partial}{\partial t}\left(\frac{a}{\Gamma}
    \right)\right]\ ,\nonumber\\
\end{eqnarray}
where we have defined the auxiliary variables
\begin{subequations}\label{BGammadef}
\begin{eqnarray}
  \Gamma&\equiv&\frac{G_\Psi}{G}=\frac{1}{1+f_R}\left(
    \frac{3+2Q}{3+3Q}\right)\ ,\label{Gammadef}\\
  B&\equiv&a\left[\frac{\partial}{\partial t}\left(\frac{a}
    {\cal H}\right)\right]^{-1}=\frac{2(1+f_R){\cal H}^2}{8\pi
    Ga^2\rho_m+a^2\partial_t(\dot f_R/a^2)}
    \label{Bdef}\ .\qquad
\end{eqnarray}
\end{subequations}

General relativity with a cosmological constant corresponds to the
case $f=2\Lambda$, $f_R=0$, and $\gamma=\Gamma=1$, yielding $U=V=0$.
Thus, $\zeta$ is conserved even on sub-horizon scales in a
$\Lambda$CDM universe.  However, this is no longer true if
$f_R\ne0$.  Two distinct effects modify the curvature perturbation.
First, the $1+f_R$ factor in (\ref{fPoisson}) modifies the evolution
on sub-horizon scales.  In practice, this effect is small if
$|f_R|\ll1$, as is favored by galactic structure considerations
\cite{HuSawicki07a}.  In this case, the background expansion history
is nearly identical to GR with a cosmological constant and gravity
is significantly modified only at wavelengths approaching the
scalaron Compton wavelength, where $Q\sim1$.  For long wavelengths
such that $Q\ll1$, $\gamma\approx1-\frac{2}{3}Q$ and the corrections
introduced to eq.\ (\ref{consistent})--(\ref{gammadef}) by scalaron
dynamics are $O(k^2)$.  The treatment given in the preceding
sections remains valid for $|f_R|\ll1$ and $Q\ll1$.  However, this
limit corresponds to general relativity. Unfortunately, the
treatment presented in Section \ref{sec:long}, which was based on a
scale-invariant modification of gravity, breaks down just where
$f(R)$ theories begin to deviate significantly from GR.

The $f(R)$ models generically have $\gamma-1\propto k^2$ for
sub-horizon wavelengths longer than the scalaron Compton wavelength.
These models have a scale-dependent modification of gravity.
Although we cannot use the results of Section \ref{sec:long} to
describe them, it is still possible to parameterize scale-dependent
modified gravity models so as to obtain useful results for the
sub-horizon growth of large-scale structure.  A simple
parameterization inspired by $f(R)$ theories is presented in the
next section.

\section{Scale-dependent modified gravity}
\label{sec:scaledep}

For a wide class of theories, modified gravity leads generically to
a Poisson equation with variable gravitational coupling.  In the
scale-invariant modifications of Section \ref{sec:long}, the Newton
constant is replaced by the time-varying $G_\Phi(t)$ which follows
from the scale-invariant potential ratio $\gamma(t)$.  In
scale-dependent modified gravity theories, on the other hand,
$G_\Phi(k,t)$ and $\gamma(k,t)=G_\Psi/G_\Phi$ are functions of
length scale as well as time, and there is no simple relation
between them. Thus, more parameters are needed to characterize such
theories \cite{Amin07}.

Despite their generality, $f(R)$ theories with $f_R\ll1$ have, for a
wide range of sub-horizon length scales, a very simple form for
$G_\Phi$ and $\gamma$ given by eq. (\ref{fPoisson}).  To arrive at a
simple phenomenological model we simplify the time dependence as
follows:
\begin{equation}\label{scaledepMG}
  \frac{G_\Phi}{G}=\frac{1+\alpha_1k^2a^s}{1+\alpha_2k^2a^s}\ ,\ \
  \gamma=\frac{1+\beta_1k^2a^s}{1+\beta_2k^2a^s}\ .
\end{equation}
We assume that these relationships hold only in the linear regime of
cosmological density perturbations, and that $G_\Phi/G\to1$ and
$\gamma\to1$ on solar system scales.  We also require GR to hold at
early times, implying $s>0$. Eq.\ (\ref{scaledepMG}) describes
$f(R)$ theories with $|f_R|\ll1$ if
$\alpha_1=\frac{4}{3}\alpha_2=2\beta_1=\beta_2=4f_{RR}/a^{2+s}$. As
a simple post-Newtonian model we will now assume that
$(\alpha_1,\alpha_2,\beta_1,\beta_2)$ are arbitrary constants with
units of length squared.  In order to ensure that $G_\Phi/G$ and $\gamma$ are finite for all $k$, we require $\alpha_2$ and $\beta_2$ to be non-negative. Moreover, we need $G_\Phi > 0$ in order to ensure that gravity is attractive. Hence, $\alpha_1$ must be non-negative as well. This scale-dependent parameterization
has a different dependence on length scale than that of Amin et al.\
\cite{Amin07}.  It is chosen to reproduce the scale-dependence of
$f(R)$ theories.  For some modified gravity theories $\gamma=1$,
e.g., Einstein plus Yukawa gravity.  For this model $G_\Phi/G$ in
eq.\ (\ref{scaledepMG}) is multiplied by an overall factor
$\alpha_2/\alpha_1$ \cite{Dore07} so that the deviation from
Einstein gravity shows up only at large distances.

The class of theories considered here has at least three physical
length scales: the horizon scale $a/{\cal H}$, the transition scale
$a^{1+s/2}\sqrt{\alpha_1}$ where gravity changes strength (for simplicity,
we consider models where the $\alpha_i$ and $\beta_i$ are all of
comparable magnitude), and the nonlinear length scale for structure
formation (e.g., approximately 10 Mpc today).  If $a^{1+s/2}\sqrt{\alpha_i}$
and $a^{1+s/2}\sqrt{\beta_i}$ are smaller than the nonlinear scale, then for
purposes of large-scale structure formation, gravity is adequately
described by GR. The parameterization of eq.\ (\ref{scaledepMG})
applies only to intermediate length scales between the horizon scale
and the smaller transition scale $a^{1+s/2}\sqrt{\alpha_1}$. However, it
implies that for long wavelengths and at early times, gravity
reduces to GR (with constant gravitational coupling). This
assumption can be relaxed at the cost of introducing additional
parameters, which seems premature given the difficulty of measuring
any post-Newtonian parameters. Also, for wavelengths short compared
with $a^{1+s/2}\sqrt{\alpha_i}$ and $a^{1+s/2}\sqrt{\beta_i}$ but large compared
with the nonlinear scale, the gravitational couplings are constant
but differ from GR, e.g., $\gamma=\frac{1}{2}$ for $f(R)$.

From the perspective of model testing, scale-dependent modified
gravity is much more complicated than the scale-independent case
considered in Sect. \ref{sec:long}.  The models have four
dimensional parameters plus an exponent giving the time dependence.
However, the situation is not so bleak, because structure formation
depends only on $G_\Phi(k,t)$ and not on $\gamma(k,t)$.  In
particular, matter density perturbations on scales larger than the
Jeans length and smaller than the Hubble length follow from
integration of
\begin{equation}\label{pertevol}
  \ddot\delta+{\cal H}\dot\delta=4\pi G_\Phi(k,t)a^2\rho_m\delta\ .
\end{equation}
At early times, $G_\Phi\to G$ and $\delta$ evolves as in the GR
solution until the scale-dependent terms in eq.\ (\ref{scaledepMG})
become important.  The density transfer function $D(k,t)$ given by
eq.\ (\ref{Dfactor}) is now scale-dependent at late times, implying
a change in the shape of the matter power spectrum.  It is easy to
see that the transfer function can depend only on the dimensionless
variables $(k\sqrt{\alpha_1},\alpha_1/\alpha_2,a)$.

\begin{figure}[t]
  \begin{center}
    \includegraphics[height = 70mm, width=90mm]{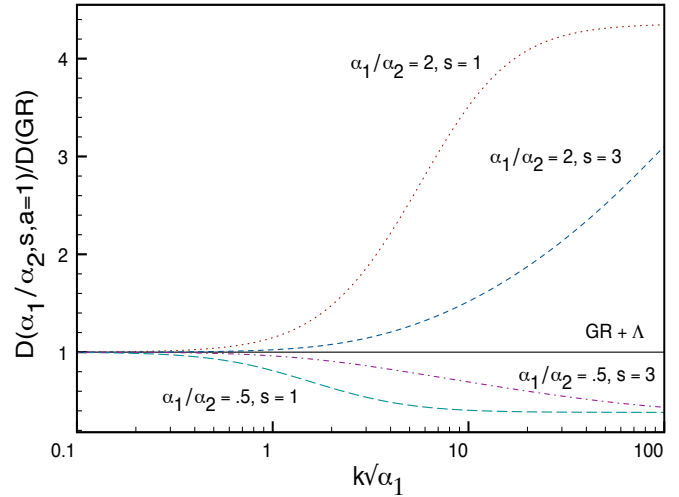}
  \end{center}
\caption{The ratio of the density transfer function to the predicted
GR case, plotted at $a=1$ versus dimensionless wavenumber
$k\sqrt{\alpha_1}$ for two values of $\alpha_1/\alpha_2$ and two
values of $s$.  For $a=0.5$ the transfer functions are almost
identical, except that $|D/D(GR)-1|$ is reduced by approximately
13\% for $\alpha_1/\alpha_2=0.5$ and 30\% for
$\alpha_1/\alpha_2=2$.} \label{fig:fundep}
\end{figure}

The most interesting new feature of scale-dependent modified gravity
is the change of shape of the matter density transfer function.
Figure \ref{fig:fundep} shows $D(k,a=1)$ normalized to the GR
result, obtained by numerically integrating eq.\ (\ref{pertevol})
with initial conditions given by the GR result ${\cal
H}^2D(a)\to\frac{2}{3}F= \frac{2}{5}$ for a matter-dominated
universe at $a=0.03$.  As expected, at large length scales
($k\sqrt{\alpha_1}\ll1$) the results converge to the GR limit.  At
short length scales, gravity is weaker than GR if
$\alpha_1/\alpha_2<1$, leading to reduced growth; the growth is
enhanced for $\alpha_1/\alpha_2>1$. Thus, scale-dependent gravity
changes the shape of the matter power spectrum \cite{Starobinsky07}.  Ultimately,
measuring this scale-dependence (and doing so at several redshifts)
can constrain scale-dependent modified gravity theories.  However,
the interpretation of the galaxy power spectrum shape is complicated
by scale-dependent biased galaxy formation and by the dark-matter-dependent
transfer function (e.g., the
neutrino fraction).  Thus, while the linear growth of structure
offers a potentially powerful test of GR versus modified gravity, it
must be combined with other tests.

\begin{figure}[t]
  \begin{center}
    \includegraphics[height = 70mm, width=90mm]{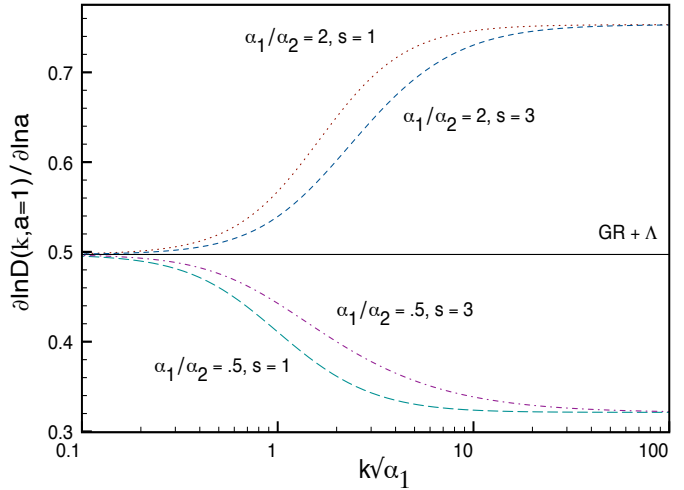}
  \end{center}
\caption{Scale-dependence of the logarithmic density perturbation
growth rate $\partial\ln D/\partial\ln a$ at $a=1$ for
scale-dependent modified gravity models.} \label{fig:S}
\end{figure}

The second function characterizing scale-dependent modified gravity,
$\gamma(k,t)$, is (in our analysis, which assumes no particular
Lagrangian) unrelated to $G_\Phi(k,t)$.  This function is best
constrained by combining weak gravitational lensing and galaxy
clustering measurements made at the same redshift.  Care is required
because the lensing amplitude is proportional to $(1+\gamma)\Phi$
while the galaxy density is proportional to $D$ and also depends on
biasing.  Galaxy peculiar velocity measurements could be used, in
principle, to reduce or ideally eliminate the dependence on biasing
\cite{ZLBD07}. However, one must be careful not to assume the
velocity-density relation obtained in GR. The continuity equation
gives
\begin{equation}\label{velden}
  {\bf v}=-\frac{i{\bf k}}{k^2}\frac{\partial\ln D}{\partial\ln a}
    {\cal H}\delta\ ,
\end{equation}
where the logarithmic growth rate $\partial\ln D/\partial\ln a$ is
now scale-dependent, as shown in Figure \ref{fig:S}.  As in the case
of galaxy clustering, measurement of this effect is contingent upon
knowing the composition of dark matter (hot dark matter has a
free-streaming scale, and its abundance determines the suppression
of growth at small scales) and correcting for any velocity bias.

The greater freedom allowed by scale-dependent modified gravity
models, and the fact that astrophysics (biased galaxy formation and
dark matter dynamics) may also introduce scale-dependence into
transfer functions, makes it challenging to test GR using growth of
structure and weak gravitational lensing.  It is likely that a
combination of galaxy clustering, peculiar velocities, and weak
lensing will be needed to obtain strong constraints on
scale-dependent modified gravity theories.

\section{Modified gravity versus shear stress}
\label{sec:shear}

A difference between the two longitudinal potentials $\Phi$ and
$\Psi$ need not signal modified gravity; it might arise from shear
stress \cite{Amendola07,Mota07}. For scalar mode fluctuations, the
shear stress is fully characterized by a scalar potential $\pi$,
such that the spatial stress tensor components are
\begin{equation}\label{shearstress}
  T^i_{\ \,j}=p\delta^i_{\ \,j}+\frac{3}{2}(\bar\rho+\bar p)\left(
    \nabla^i\nabla_j-\frac{1}{3}\delta^i_{\ \,j}\Delta\right)\pi
\end{equation}
where $(\bar\rho+\bar p)$ is the background enthalpy and
$\Delta=\nabla^i\nabla_i$. In linearized GR, one of the Einstein
field equations yields
\begin{equation}\label{GRshear}
  \Psi-\Phi=12\pi Ga^2(\bar\rho+\bar p)\pi\ .
\end{equation}
All of the results obtained in Sections \ref{sec:long} and
\ref{sec:observe} for modified gravity apply equally to GR with
shear stress if $\gamma$ is replaced by $12\pi Ga^2(\bar\rho+\bar
p)\pi/\Phi$.

In standard cosmology, the only significant source of shear stress
is relativistic neutrinos after neutrino decoupling during the
radiation-dominated era.  For long wavelengths during the
radiation-dominated era, neutrino shear stress gives
\cite{MaBert95}
\begin{equation}\label{nushear}
  \gamma-1=\frac{2}{5}\left(\frac{\rho_\nu+p_\nu}{\bar\rho+\bar p}
    \right)\ .
\end{equation}
During the matter-dominated era, $\gamma-1\propto a^{-1}$ and shear
stress is unimportant at late times in the $\Lambda$CDM model.  It
is also unimportant in simple quintessence models because shear
stress vanishes for linear perturbations of a minimally-coupled
scalar field.

Shear stress might nonetheless be important if cosmic acceleration
is driven by an imperfect fluid.  Without specifying the dynamics of
this fluid, few constraints can be placed on $\pi$.  One possible
bound comes from the dominant energy condition, which states that
each of the eigenvalues of $T^i_{\ \,j}$ must be smaller in absolute
value than the energy density.  If this condition holds, then eqs.\
(\ref{shearstress}) and (\ref{GRshear}) can be combined to give
rather weak bounds on $\Psi/\Phi-1$.

Additional constraints follow from the initial-value constraints of
GR and energy-momentum conservation, which for a spatially flat
background become \cite{Bertschinger06}
\begin{subequations}\label{ivcon}
\begin{eqnarray}
  -k^2\Psi&=&4\pi Ga^2(\bar\rho+\bar p)(\delta+3{\cal H}u)\ ,
    \qquad\label{Energycon}\\
  \dot\Psi+{\cal H}\Phi&=&4\pi Ga^2(\bar\rho+\bar p)u\ ,
    \label{Momcon}\\
  \dot\delta+3{\cal H}\sigma&=&3\dot\Psi-k^2u\ ,
    \label{Econs}\\
  \dot u+(1-3c_s^2){\cal H}u&=&c_s^2\delta+\sigma+\Phi-k^2\pi\ .
    \label{Momcons}
\end{eqnarray}
\end{subequations}
The first of these equations is the usual Poisson equation in
conformal Newtonian gauge. The density and velocity potential
perturbations are defined from the energy-momentum tensor components
by $T^0_{\ \,0}=-\bar\rho-(\bar\rho+\bar p)\delta$, $T^0_{\ \,i}=
-(\bar\rho+\bar p)\nabla_iu$, while the entropy perturbation is
defined by $\sigma\equiv(\delta p-c_s^2\delta\rho)/(\bar\rho+\bar
p)$ with $c_s^2=d\bar p/d\bar\rho$.  For a single perfect fluid like
cold dark matter before its trajectories intersect, $\sigma=0$.
However, in general $\sigma\ne0$ for a multi-component fluid, e.g.
dark matter and a non-constant dark energy.

The freedom introduced by entropy and shear stress perturbations is,
unfortunately, sufficient in principle to reproduce any observations
of large-scale structure and gravitational lensing.  Consider, for
example, perfect measurements of galaxy peculiar velocities
everywhere and at all times assuming that galaxies exactly trace
cold dark matter. Then, eq.\ (\ref{Momcons}) with
$c_s^2=\sigma=\pi=0$ for CDM yields $\Phi({\bf k},t)$. Assume
furthermore that complete and ideal gravitational lensing
measurements are available to yield $\Phi({\bf k},t)+\Psi({\bf
k},t)$.  Now, the GR initial-value constraints
(\ref{Energycon})--(\ref{Momcon}) suffice to yield $\delta({\bf
k},t)$ and $u({\bf k},t)$ for the multi-component fluid of dark
matter and dark energy.  Requiring this fluid to obey
energy-momentum conservation (\ref{Econs})--(\ref{Momcons}) yields
$\sigma$ and $\pi$.  In short, perfect measurements of $\Phi$ and
$\Psi$ can be used to determine $\sigma$ and $\pi$ of the combined
fluid of dark matter and dark energy.  When combined with
measurements of the dark matter density and velocity fields
(assuming galaxies trace dark matter), one can, in principle,
determine the energy-momentum tensor components for dark energy. One can think of this energy-momentum tensor for dark energy as the difference between the Einstein tensor for the observed metric and the energy-momentum tensor of the ordinary matter \cite{HuSawicki07b}.

This approach can be used, in principle, to determine the dark
energy entropy and shear stress needed to explain any observations
of large-scale structure (including peculiar velocities) and weak
lensing --- even if there is no dark energy, but instead gravity is
modified.  In effect, the two observable metric fields $\Phi$ and
$\Psi$ can be exchanged for either $\sigma$ and $\pi$ (GR with
exotic dark energy) or $G_\Phi$ and $\gamma$ (modified gravity). 

Although the initial-value constraints of GR can be used to
determine the properties of dark energy, they cannot be used to
model modified gravity.  As we have seen in previous sections,
modified gravity leads generically to a modified Poisson equation
with a variable gravitational coupling $G_\Phi(k,t)$.  In order not
to break local Lorentz invariance by the selection of a preferred
frame, the other components of the Einstein equation should also be
modified.  For example, in $f(R)$ theories the left-hand side of
(\ref{Momcon}) is multiplied by $(1+f_R)$ while the right-hand side
acquires an extra term $\frac{1}{2}\dot f_R$.  Neglecting these
modifications leads to violations of eq.\ (\ref{consistent}) on
large scales.  For example, Caldwell et al.\ \cite{Caldwell07}
modeled the dynamics of modified gravity with three recipes that
assume the validity of some combination of eqs.\ (\ref{Energycon})
and (\ref{Momcon}).  While recipes R1 and R2 satisfy the conservation of $\zeta$, recipe R3 does not. As a result, it leads to a different prediction for the gravitational potentials and therefore the ISW effect.


\section{Discussion}
\label{sec:discuss}

Parameterizing modified gravity theories in cosmology is much more
difficult than parameterizing post-Newtonian gravity in the solar
system because the gravitational potentials $\Phi$ and $\Psi$ in the
GR limit are not static Coulomb potentials in cosmology.  In GR, on
scales larger than the Jeans or nonlinear length scales, $\Psi=\Phi$
factorizes into a product of a time-dependent growth function and a
spatially-varying curvature perturbation.  If this factorization
persists in modified gravity theories, then a simple post-Newtonian
parameterization can be obtained.  This is the approach we
introduced in Section \ref{sec:long}.  It has the practical virtue
of yielding easily calculated predictions for all observables in the
linear regime, including the growth of matter clustering, peculiar
velocities, microwave background anisotropy, and weak gravitational
lensing.

Introducing the Eddington parameterization $\Psi=\gamma\Phi$, where
$\gamma$ depends on time but not on space, we showed that structure
grows faster (gravity is stronger) in models with $\gamma<1$.
However, the shape of the transfer functions (i.e., their dependence
on spatial wavenumber) for these scale-independent models is
unchanged compared with GR.

Unfortunately, realistic modified gravity theories such as the
$f(R)$ models have scale-dependent effects and can no longer be
described by only one post-Newtonian quantity $\gamma$.  Instead,
the strength of gravity, described by $G_\Phi$ in the Poisson
equation, can vary independently of $\gamma$.  Even so, because
galaxy clustering and dynamics depends on $G_\Phi$ but not on
$\gamma$, while weak lensing depends on $\gamma$, observations
could, in principle, measure modified gravity parameters assuming
there is no dark energy.

If dark energy is complex enough to require two additional fields to
characterize its stress tensor (e.g., shear stress potential and
entropy), then it appears that there is enough information in the
dark energy model to account for any $\Phi$ and $\Psi$ without
modifying gravity.  One cannot prove gravity is modified unless one
can account for all significant contributions to the stress-energy
tensor.

Thus, our hope to describe all modified gravity models with two
parameters, yielding predictions measurably different from all dark
energy models, has not been realized.  Distinguishing modified
gravity from dark energy will require making additional assumptions.

Nevertheless, the $(\beta,s)$ parameterization of scale-independent
modified gravity presented in Section \ref{sec:long}, and the
$(\alpha_1,\alpha_2,\beta_1,\beta_2)$ parameterization of
scale-dependent modified gravity models presented in Section
\ref{sec:scaledep}, are still useful for characterizing
observational data.  If measurements of galaxy clustering, peculiar
velocities, and weak lensing are all consistent with $\beta=0$ and
$\alpha_1=\alpha_2=\beta_1=\beta_2=0$, for example, then modified
gravity and exotic dark energy models can both be excluded.  If
measurements require nonzero parameters, however, dark energy and
modified gravity remain viable explanations until additional
assumptions are made to distinguish them,e.g.\ restriction of the
Lagrangian to a particular form.

A generic prediction of modified gravity theories in cosmology is
that the gravitational coupling $G_\Phi$ in the Poisson equation
should vary with time and with length scale.  Departures from GR
could be important not only in the linear regime of cosmological
perturbations but perhaps also in the nonlinear regime (albeit on
scales much larger than the solar system).  Nonlinear effects may
allow modified gravity to be distinguished from exotic dark energy,
assuming that the dark energy fluctuations are small.  For this
reason it would be valuable to perform N-body simulations of
structure formation using variable $G_\Phi$, extending previous work
\cite{Nbody} to the scale-independent and scale-dependent modified
gravity models discussed in the current paper.

\acknowledgments

We thank Richard Gott for helpful comments and Scott Tremaine for
the hospitality of the Institute for Advanced Study.  This work was
supported by the Kavli Foundation, by a Guggenheim Fellowship to EB,
and by NSF grant AST-0708501.

\end{document}